\documentclass[conference]{IEEEtran}
\IEEEoverridecommandlockouts
\usepackage{cite}
\usepackage{amsmath,amssymb,amsfonts}
\usepackage{algorithmic}
\usepackage{graphicx}
\usepackage{graphics}
\usepackage{textcomp}
\usepackage{xcolor}
\usepackage{multirow}
\usepackage[super]{nth}
\def\BibTeX{{\rm B\kern-.05em{\sc i\kern-.025em b}\kern-.08em
    T\kern-.1667em\lower.7ex\hbox{E}\kern-.125emX}}
\begin{document}
\makeatletter 
\newcommand{\linebreakand}{%
  \end{@IEEEauthorhalign}
  \hfill\mbox{}\par
  \mbox{}\hfill\begin{@IEEEauthorhalign}
}
\makeatother 

\title{Influence of Video Dynamics on EEG-based Single-Trial Video Target Surveillance System

\footnote{
\thanks{This work was supported by the Agency For Defense Development Grant Funded by the Korean Government (UI233002TD).}}
}

\author{\IEEEauthorblockN{Heon-Gyu Kwak}
\IEEEauthorblockA{\textit{Dept. of Artificial Intelligence} \\
\textit{Korea University} \\
Seoul, Republic of Korea \\
hg\_kwak@korea.ac.kr}
\and
\IEEEauthorblockN{Sung-Jin Kim}
\IEEEauthorblockA{\textit{Dept. of Artificial Intelligence} \\
\textit{Korea University} \\
Seoul, Republic of Korea \\
s\_j\_kim@korea.ac.kr}
\and 
\IEEEauthorblockN{Hyeon-Taek Han}
\IEEEauthorblockA{\textit{Dept. of Artificial Intelligence} \\
\textit{Korea University} \\
Seoul, Republic of Korea \\
ht\_han@korea.ac.kr}
\linebreakand

\IEEEauthorblockN{Ji-Hoon Jeong}
\IEEEauthorblockA{\textit{Dept. of Computer Science} \\
\textit{Chungbuk National University} \\
Cheongju, Republic of Korea \\
jh.jeong@chungbuk.ac.kr}

\and
\IEEEauthorblockN{Seong-Whan Lee}
\IEEEauthorblockA{\textit{Dept. of Artificial Intelligence} \\
\textit{Korea University} \\
Seoul, Republic of Korea \\
sw.lee@korea.ac.kr}
}
\maketitle 

\begin{abstract}
Target detection models are one of the widely used deep learning-based applications for reducing human efforts on video surveillance and patrol. However, the application of conventional computer vision-based target detection models in military usage can result in limited performance, due to the lack of sample data of hostile targets. In this paper, we present the possibility of the electroencephalography-based video target detection model, which could be applied as a supportive module of the military video surveillance system. The proposed framework and detection model showed prospective performance achieving a mean macro $F_{\beta}$ of 0.6522 with asynchronous real-time data from five subjects, in a certain video stimulus, but not on some video stimuli. By analyzing the results of experiments using each video stimulus, we present the factors that would affect the performance of electroencephalography-based video target detection models.

\end{abstract}

\begin{small}
\end{small}

\section{INTRODUCTION}

Over the past few years, deep learning has been improved in revolutionary advances across various domains, with video recognition and analysis standing out as one of its most promising applications. Specifically, video target detection plays an important role in urban safety surveillance, traffic management, and various commercial applications \cite{voulodimos2018deep, sreenu2019intelligent, dargan2020survey}. However, in specialized fields such as military surveillance and patrol, conventional computer vision (CV)-based detection methodologies exhibit limitations due to the limited number of hostile target samples \cite{zhang2023uav}.

To address this challenge approaches leveraging human perception and cognition using brain computer-interface (BCI) techniques on the target detection task have been proposed recently \cite{song2020asynchronous, zhang2023uav}, as the effectiveness of BCI techniques has been proved \cite{kim2015abstract, thung2018conversion, supratak2017deepsleepnet}. Among various neuroimaging methods, electroencephalogram (EEG) would be a reasonable method for implementing the target detection system using brain signals, due to its advantages in mobility, low cost, and high time-resolution \cite{lee2019possible}, as many studies presented in various fields of real-time BCI applications \cite{soleymani2015analysis, lee2019towards, lee2020neural, perslev2021u,  lee2019comparative, vidyaratne2017real, mane2021fbcnet, bang2021spatio}.

In this paper, we propose the framework for implementing a video target detection model based on EEG, which can possibly be the supportive module to cooperate with CV based target detection system. Furthermore, by analyzing the trained model, we present insights in constructing the framework for EEG-based video target detection.

\begin{figure*}[ht]
    \centering
    \includegraphics[width=18cm]{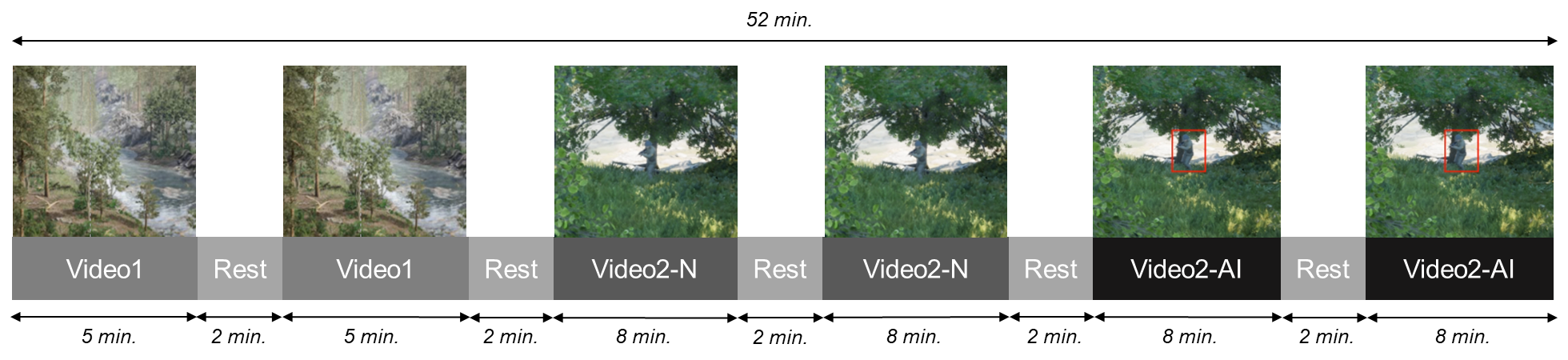}
    \caption{An overview of the EEG data acquisition process. A subject first watches `Video1' twice, and then watches `Video2-N' and `Video2-AI' twice in turn. Each video clip is eight minutes long. The subject takes rests for two minutes after a session ends. Each subject proceeds four sessions of EEG acquisition, taking 52 minutes in total.}
    \label{fig:fig1}
\end{figure*}

\begin{figure}[ht]
    \centering
    \includegraphics[width=6cm]{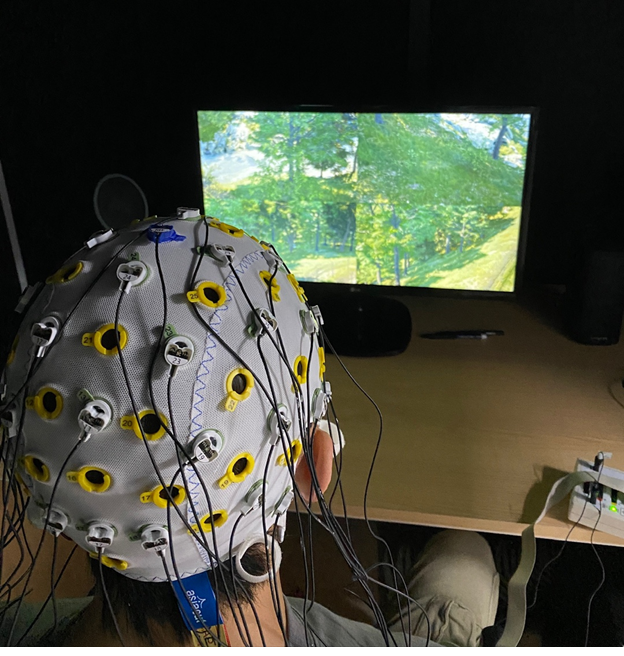}
    \caption{An example of an EEG data acquiring session with surveillance video clips. Subjects were instructed to concentrate on the presented video clip to detect appearing targets. EEG signals were recorded using 32-channel electrodes, in the soundproof experimental booth with lights off.}
    \label{fig:fig2}
\end{figure}

\section{MATERIALS AND METHODS}

\subsection{Subjects}
A total of 13 healthy male and female subjects participated in the EEG data acquiring task. All subjects were aged between 20 and 40 years old and had no neurophysiological or mental disorders. Four of the subjects participated in the offline EEG data acquiring process, which is used for training dataset of the target detection model. The other nine subjects participated in the online testing experiments which are for evaluating the model performance in a real-time asynchronous environment. 

\begin{table}[]
\centering
\caption{Differences in video stimuli}
\renewcommand{\arraystretch}{1.2}
\begin{tabular}{cccc}
\hline
\multirow{2}{*}{Dynamics} & \multicolumn{3}{c}{Name}                                     \\ \cline{2-4} 
                          & \multicolumn{1}{c}{Video1} & \multicolumn{1}{c}{Video2-N} & Video2-AI \\ \hline
Camera rotation           & \multicolumn{1}{c}{x}      & \multicolumn{1}{c}{o}        & o         \\ \hline
Weather changing          & \multicolumn{1}{c}{x}      & \multicolumn{1}{c}{o}        & o         \\ \hline
Target bounding box       & \multicolumn{1}{c}{x}      & \multicolumn{1}{c}{x}        & o         \\ \hline
\end{tabular}
\end{table}

\subsection{Video Target Stimuli}
Three video clips `Video1', `Video2-N', and `Video2-AI' are used as stimuli for acquiring EEG signals from subjects. Each video clip is simulated video surveillance footage, assumed to be recorded in a mountain forest area. Each of these clips is divided into quarters, which shows various scenes of surveillance cameras in different settings of angles and positions. `Video1' is 300 seconds long, while 'Video2-N' and 'Video2-AI' are 480 seconds long. 

For Video2-N' and 'Video2-AI', surveillance cameras rotate for three seconds to monitor other sites every five seconds periodically, in turn. Moreover, the weather of the surveillance location changes into foggy or rainy as time goes by.

In each video clip, three types of targets `deer', `wild boar', and `soldier' suddenly appear and last for one second several times, in a specific time point and location. Deer and wild boars are non-hostile targets which can be classified as error-targets, and soldiers are hostile targets representing true-target. 

Two video clips `Video2-N' and `Video2-AI' are essentially identical except for one difference, `Video2-AI' shows a red bounding box that guides the locations of targets when they appear, whereas `Video2-N' does not. `Video1' has different timing, location, and the number of times in target appearance with `Video2-N' and `Video2-AI'. Table I shows the differences between video clips.

\subsection{EEG Data Acquisition Process}
Each of the four subjects was instructed to concentrate on video clips to detect each target while equipping the EEG acquiring devices. EEG signals were recorded at a 250 Hz of sampling rate, with the BrainVision Recorder software provided by BrainProduct (GmbH, Germany), using 32 Ag/AgCL electrodes placed in 10-20 international systems. During the acquisition, the impedance of each electrode was maintained under 10 k$\Omega$ for excluding noises caused by poor contactivity of electrodes. To minimize artifacts, the recording proceeded in a soundproof experimental booth while the lights were off, and the subjects were asked to minimize facial and body movements. Each subject watched all of the video clips twice each, six sessions of watching in total, and had two minutes of resting time when a session ended. Fig. 1 represents the overview of the configuration of the EEG data acquisition process, and Fig. 2 shows the example of an EEG acquiring session.

\subsection{Model Training}
We hypothesized that there would be general EEG patterns related to the target detection across subjects. Whereas, bio-signals like EEG have high variations in patterns between individuals \cite{lin2012generalized, suk2014predicting, li2018exploring, kim2019subject}, even if they perform the same task or are in the same condition, resulting from unique neurological and physiological characteristics of individuals. To consider both of these factors, we pre-trained the model using all of the subject's EEG data, and fine-tuned model parameters with calibration when testing.

\subsubsection{Model Architecture}
We modified the DeepConvNet\cite{schirrmeister2017deep} into hierarchical architecture, which classifies non-target and targets (true and error) as binary classes first, and re-classifies true- and error-target after it. The main reason for two-stage classification is that we hypothesized that classifying non-target and target is a relatively easy task compared to classifying true-target and error-target, thus, separating learning parameters would result in more stable performance.

\subsubsection{Training Objective}
We defined the target detection problem of this study as a highly imbalanced 3-class classification task, with the classes of 0 (non-target), 1 (true-target), and 2 (error-target). Therefore, we used multi-class cross-entropy loss for training the model. The multi-class cross-entropy loss is defined as follows:

\begin{align}
&\mathcal{L}(\hat{y}, y)=-{\sum^2_{k=0}}y^{c} \log \hat{y}^{c}, 
\end{align} where $y$ is the true label of the input data, $\hat{y}$ is the predicted label of the input data, and $c$ is the index of classes of the input dataset.

\subsubsection{Target Label Assigning}
For the data of an EEG acquisition session, the total number of data time points is sampling rate times video length. For example, the EEG data acquired with `Video1' has 75,000 time points ($T=[t_1, t_2, t_3, \dots, t_{74999}, t_{75000}])$. In this set of time points $T$, a target $c$ appears on a certain time point  $t_x$ and lasts for one second. Thus, the label $y^c$ would be assigned to the set of time points $[t_{x}, t_{x+1}, \dots, t_{x+249}, t_{x+250}]$, therefore, the set of labels $Y_{t_x}^{c}=[y^{c}_{x}, y^{c}_{x+1}, \dots, y^{c}_{x+249}, y^{c}_{x+250}]$, where $y^c_x=[t_x, y^c]$. For any other time points that are not assigned by class 1 or 2, we assigned 0 as a label.

\subsubsection{Data Augmentation}
Compared to the total duration of the video clip, the duration of appearances for each target is enormously small. Even each type of class appears 30 times each on the `Video2-N', the class ratio between three classes is non-target (87.5 \%), true-target (6.25 \%), and error-target (6.25 \%). This kind of highly imbalanced dataset would drive the model to learn the biased features of the majority class and lower the generalized performance of the model. To alleviate this problem, we augmented data of the minority class using the sliding window method which is known as effective data augmentation method for EEG data \cite{lashgari2020data}. We used the window size with 250 time points with a stride of 25 time points, to overlap the input data in 0.1 seconds of interval. As a result, the volume of the minority data is augmented to ten times larger than its original volume.  

\subsubsection{Training Hyperparameters}
We optimized the trainable parameters of the model using Adam optimizer \cite{kingma2017adam}. Hyperparameters used for the training were batch size of 128, learning rate of 0.001, weight decay of 0.0001, dropout rate of 0.1, and 100 training epochs.

\subsection{Evaluation Metric}
To consider the imbalance of each class, The performances of the model were evaluated with macro $F_{\beta}$, which can be defined as

\begin{align}
    &\textit{recall} = {\textit{true positive} \over \textit{true positive}+\textit{false negative}}, \\
    &\textit{precision} = {\textit{true positive} \over \textit{true positive}+\textit{false positive}}, \\
    &\textit{F}_{\beta} = (1+{\beta}^2) \cdot {{\textit{recall} \cdot \textit{precision}} \over {(\beta^2 \cdot \textit{recall})+\textit{precision}}}, \\ 
    &\textit{macro F}_{\beta} = {1 \over N} {\sum}^{N}_{c=0} F_{\beta}^c,
\end{align}
where $N$ is the number of classes, $F_{\beta}^c$ is the $F_{\beta}$ of the class $c$, and $\beta$ is the parameter to adjust weights of recall and precision. In this study, $N$ is three because we defined the problem as a 3-class classification task, and we set $\beta$ as two to give larger weight on the recall.

\begin{figure*}[t]
    \centering
    \includegraphics[width=18cm]{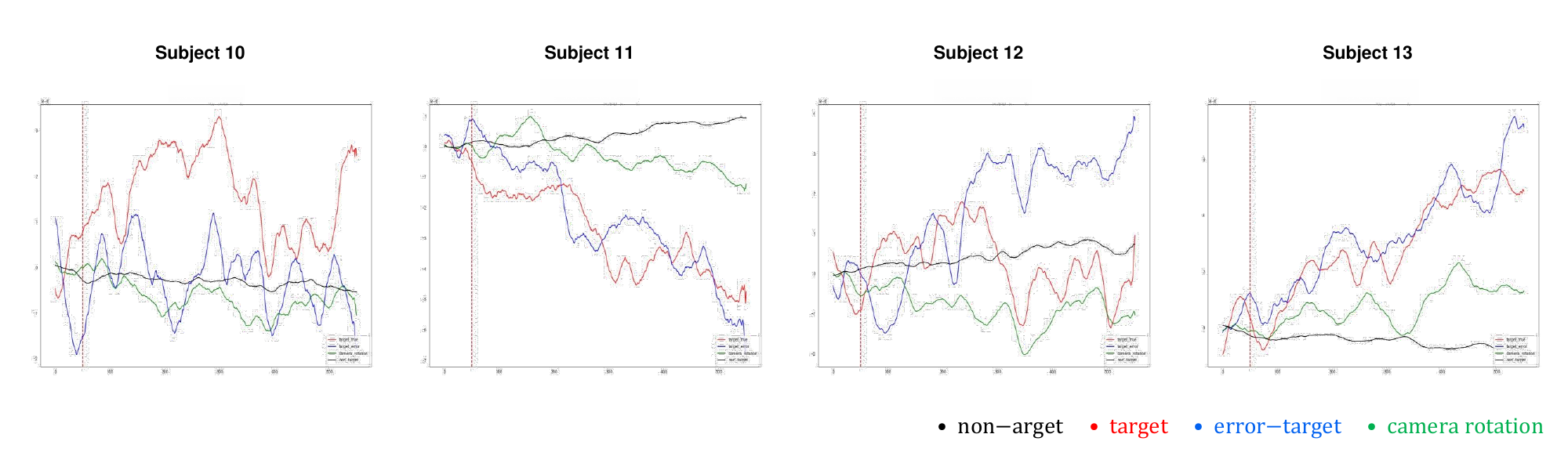}
    \caption{Grand average of ERPs for subject 10, 11, 12, and 13 by class, with EEG data of `Video2-N'. The figure contains the ERPs for three seconds of duration after the target appearance. The red vertical line indicates the time point of the target appearance. The black plot is for the ERPs of non-target, the red plot is for true-targets (enemy soldier), blue plot is for error-targets, and green plot is for camera rotations}
    \label{fig:fig3}
\end{figure*}

\section{EXPERIMENTAL SETTINGS}
We conducted experiments to evaluate the performance of the proposed EEG-based video target detection model in the online environment setting. Nine subjects who did not engage in the offline EEG data acquiring process participated in the online test session. To receive EEG signals asynchronously in real-time, we leveraged the remote data access function of the BrainVision. Out of nine subjects, five subjects watched `Video1' as video stimuli, and the rest of the four subjects used `Video2-N' and `Video2-AI' as video stimuli.

\section{RESULTS AND DISCUSSIONS}
\subsection{Online Inference Performance}
In the experiments with `Video1' for five subjects, the proposed model achieved the macro $F_{\beta}$ of 0.7000, 0.5238, 0.7568, 0.5238, and 0.7568 for subject 5, 6, 7, 8, and 9 respectively, and the overall performance was 0.6522. When the model was trained with the EEG data of `Video2-N', the performance resulted in 0.1200, 0.1569, 0.0392, and 0.1967 for subject 10, 11, 12, and 13 respectively. The model trained with the EEG data from `Video2-AI' showed the performance of 0.0784, 0.1091, 0.1200, 0.0952, and 0.1007 for subject 10, 11, 12, and 13. The model trained with EEG data of `Video1` showed the highest performance, while the data of `Video2-AI' resulted in the lowest performance. Table II shows the overall performances of the experiments. 

The model trained with the EEG data of `Video1' showed promising performance in the EEG-based target detection task. In contrast, the model trained with the EEG data of `Video2-N` and `Video2-AI' showed poor performance, even if other factors that affect model performance, such as the model architecture, the number of parameters, and training hyperparameters, were set in the same condition with the model of `Video1'. For this performance gap, we analyzed the model and input EEG signals to identify any other factors that might affected the results.

\begin{table}[]
\centering
\caption{Online Inference Performance by Video Stimulus}
\renewcommand{\arraystretch}{1.2}
\begin{tabular}{c c c c}
\hline
\multicolumn{4}{c}{\textbf{Performance (macro $F_{\beta}$)}}             \\ \hline
\multirow{2}{*}{\textbf{Subject}} & \multicolumn{3}{c}{\textbf{Video Stimulus}} \\ \cline{2-4} 
                         & \textbf{Video1}   & \textbf{Video2-N}  & \textbf{Video2-AI}  \\ \hline
S5                       & 0.7000   & -         & -          \\ \hline
S6                       & 0.5238   & -         & -          \\ \hline
S7                       & 0.7568   & -         & -          \\ \hline
S8                       & 0.5238   & -         & -          \\ \hline
S9                       & 0.7568   & -         & -          \\ \hline
S10                      & -        & 0.1200    & 0.0784     \\ \hline
S11                      & -        & 0.1569    & 0.1091     \\ \hline
S12                      & -        & 0.0392    & 0.1200     \\ \hline
S13                      & -        & 0.1967    & 0.0952     \\ \hline
\textbf{Mean}                     & \textbf{0.6522}   & \textbf{0.1282}    & \textbf{0.1007}     \\ \hline
\footnotesize{S is the abbreviation of subject.}
\end{tabular}
\end{table}

\subsection{Event-related Potentials (ERPs) Analysis}
ERPs are commonly considered important EEG feature that highly influences the decision of the EEG-based target detection models \cite{song2020asynchronous, zhang2023uav}. Therefore, we calculated the grand average of all trials of ERPs from `Cz', `C3', and `C4' channels, to identify if the ERPs played an important feature of the model. Fig. 3 represents the result of the ERPs by classes, in camera rotation. In the figure, subject 11 and 13 showed better performances compared to the subject 10 and 12, but their ERPs showed less clearly distinct differences in patterns between classes. This result implies that the model did not count temporal patters of ERPs as important features, for classifying each class.

\subsection{Saliency Map Analysis}
We also analyzed the model using a saliency map \cite{adebayo2018sanity}, to visualize the importance of each EEG channel for the model inference. In Fig. 4, the area colored in deep blue represents that the model performance decreases when the EEG channel data of the area is removed, meaning the EEG channel plays an important role for the trained model in inference. The EEG channels of the red area represent the relatively lower importance of the channel compared to others. As shown in the figure, important channels for the trained model were mainly are located in central, temporal and occipital regions, which are related to the passive visual perception, like camera rotation, rather than the frontal and temporal-parietal regions which are mostly related to dicsciminative response task, such as, target detection \cite{di2019normative, menon1997combined}. 

These results would be caused by differences in video dynamics, such as camera rotation and weather changes, arousing the passive visual perception-related EEG features more than the discriminative response task-related EEG features to the subject \cite{song2021p3}. Imbalanced inductions of EEG features would lead the model to be biased on passive visual perception-related EEG features, therefore, the performance of target detection would be poor.

\begin{figure}[t]
    \centering
    \includegraphics[width=9cm]{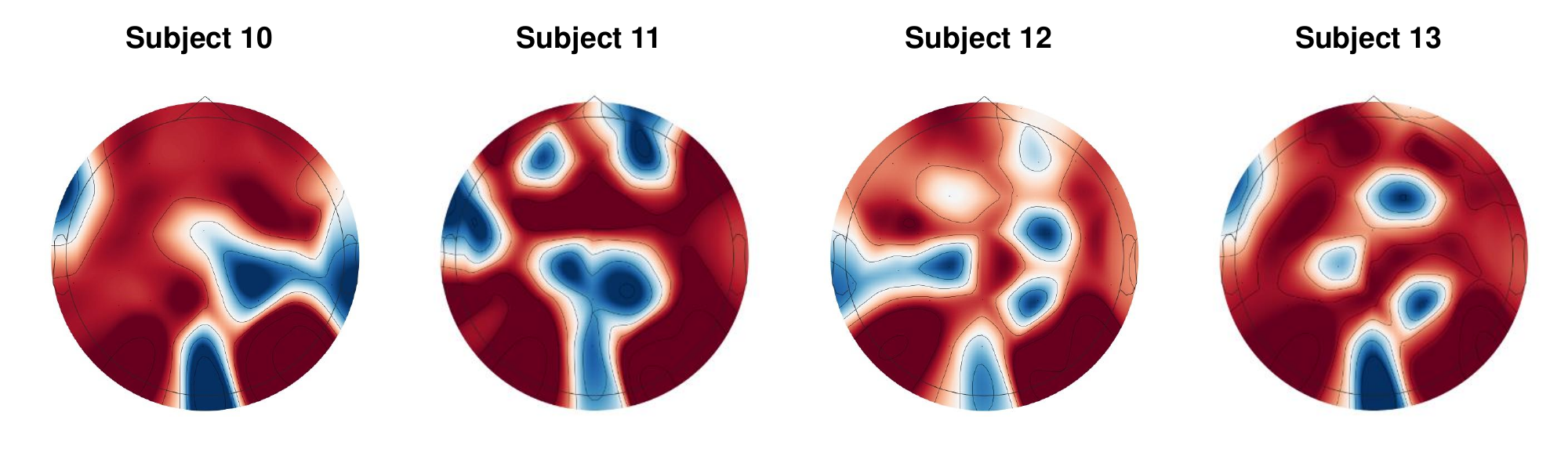}
    \caption{Channel-wise saliency map of subject 10, 11, 12, and 13 with EEG data of `Video2-AI'. Deep blue colored areas indicate that EEG channels located in the area are relatively more important than other channels in model inference. Red colored areas indicates vice versa.}
    \label{fig:fig4}
\end{figure}

\section{CONCLUSION}
In this paper, we presented the possibility of an EEG-based video target detection system in real time asynchronous environment, while showing the system would fail when video stimuli have specific characteristics, like repeated periodically video dynamics that causes passive visual perception, leading the model to train with irrelative EEG features for target detection. To implement a more stable and precise EEG-based target detection system, researchers should carefully design the experimental paradigm and protocol. We hope that this study will give insights to researchers, and help to advance the field of EEG-based target detection studies.

\bibliographystyle{IEEEtran}
\bibliography{MANUSCRIPT}

\end{document}